\documentclass[apjl]{emulateapj}
\usepackage{amssymb,graphicx}
\usepackage{epsfig}
\usepackage{psfrag}
\usepackage[usenames]{color}
\usepackage{multirow}
\usepackage{rotating}

\shorttitle{Dynamical Capture NS Mergers}
\shortauthors{East \& Pretorius}

\begin{document}

\title{Dynamical Capture Binary Neutron Star Mergers}

\author{William E.\ East and Frans Pretorius}
\affil{Department of Physics, Princeton University, Princeton, NJ 08544, USA.}

\begin{abstract}
We study dynamical capture binary neutron star mergers as may arise in dense
stellar regions such as globular clusters.  Using general-relativistic hydrodynamics,
we find that these mergers can
result in the prompt collapse to a black hole or in the formation of a hypermassive neutron star,
depending not only on the neutron star equation of state but also on impact
parameter.  We also find that these mergers can produce accretion disks of up to
a tenth of a solar mass and unbound ejected material of up to a
few percent of a solar mass.
We comment on the gravitational radiation and electromagnetic transients that these sources may produce.
\end{abstract}


\maketitle

%--------------------------------------------------------------------------
% Introduction
%--------------------------------------------------------------------------
\section{Introduction}

Merging binary neutron stars (NSs) warrant detailed study because these systems promise
to be rich sources of both gravitational and electromagnetic (EM) radiation, probing
strong-field gravity and nuclear density physics.  NS--NS 
mergers are a primary source targeted by gravitational wave (GW) 
detectors~\citep[such as LIGO;][]{LIGO}.  They are also candidates for short gamma-ray burst (SGRB) 
progenitors and several other EM counterparts~\citep{2012ApJ...746...48M,2041-8205-736-1-L21,2012arXiv1204.6242P}  
which could potentially be observed by current and upcoming wide-field
survey telescopes like PTF~\citep{2009PASP..121.1334R}, Pan-STARRS~\citep{2004SPIE.5489...11K}, 
and LSST~\citep{2009arXiv0912.0201L}.

There have been numerous studies of primordial binary NS mergers (see e.g.~\cite{faber_review}),
which will have essentially zero orbital
eccentricity when they enter the frequency band of ground-based GW detectors.
However, binaries may also form via $n$-body interactions in dense stellar regions and some fraction
of them will have sizable eccentricity at merger.
A likely environment to find such binaries is 
globular clusters (GCs) that have undergone core collapse \citep{fabian75,grindlay2006}.
In~\cite{lee2010}, it was argued that the rate of tidal capture and collision of two NSs in 
GCs (using M15 as a prototype) peaked around $z\simeq0.7$
at values of $\sim50$ yr$^{-1}$ Gpc$^{-3}$ (falling to $\sim30$ yr$^{-1}$ Gpc$^{-3}$ by $z=0$) and was 
consistent with the observed SGRB rate. 
However, this does not take into account natal kicks. A recent simulation of M15
that assumed a modest NS retention fraction of $5\%$ found $\sim1/4$ fewer NSs in the central
$0.2$ pc~\citep{2011ApJ...732...67M} compared to an earlier study that ignored kicks~\citep{dull,2003ApJ...585..598D},
implying the above rate (scaling as the number density squared) could be overestimated by
an order of magnitude.  
On the other hand, observations suggest that the NS retention fraction in some GCs can
be as large as $20\%$~\citep{2002ApJ...573..283P}, and the model of~\cite{lee2010} did not
take into account other channels that could lead to binary merger within a Hubble
time, such as Kozai resonance in a triple system~\citep{Thompson:2010dp}.

The above discussion focused on GC environments, and
similar interactions in galactic nuclei would also add to the rates~\citep{oleary,Kocsis:2011jy,Antonini2012}.
Still, it is far from certain that high eccentricity mergers occur
frequently enough to expect observation with the upcoming generation of GW detectors.
However, it is also not implausible that they do, and
as eccentric NS mergers may also produce distinguishable EM emission compared to quasi-circular
mergers, it behooves us to understand both systems from a multi-messenger
perspective.

In~\cite{ebhns_letter} and \cite{bhns_astro_paper}, black-hole--neutron-star (BH--NS) mergers
formed through dynamical capture were found to exhibit a rich variation with impact 
parameter, in some cases producing sizable disks and amounts of unbound material.
In~\cite{gold}, several eccentric NS--NS mergers were studied using a $\Gamma=2$ equation
of state (EOS) and shown to exhibit $f$-mode excitation during close encounters.  
There have also been studies of BH--NS and NS--NS collisions with Newtonian
gravity~\citep{lee2010,2012arXiv1204.6240R} showing similar variation in the outcomes.

In this Letter, we study dynamical 
capture NS--NS mergers for a 
range of impact parameters using general-relativistic hydrodynamics (GRHD).  
We also consider several different NS EOSs because of the uncertainty regarding the 
correct description of matter above nuclear densities. One of the 
important issues we address for the first time is if these mergers
can produce hypermassive neutron stars (HMNSs).  In studies of quasi-circular systems it
was found that thermal energy from the merger, as well as differential rotation, could support
long-lived HMNSs for some EOSs~\citep[e.g.,][]{PhysRevLett.107.051102} and that this would be imprinted in the GW signal and resulting disk
properties.  HMNSs with longer lifetimes can also build up significant magnetic fields which 
can power strong EM transients during the collapse to a BH~\citep{2011arXiv1112.2622L}.
For dynamical capture binaries, the amount of angular momentum, and likely 
the amount of shock heating, will be strong functions of impact parameter, suggesting
HMNS formation will be as well.

Another notable feature of dynamical capture NS--NS mergers is their potential to produce unbound nuclear material 
which will decompress and form heavy nuclei via the $r$-process
~\citep{1974ApJ...192L.145L,Rosswog:1998gc,Li:1998bw}; subsequent radioactive decay
could produce observable emission. Recent work~\citep{Fischer,Arcones} suggests 
processes like NS--NS mergers may be needed to supplement the supernovae $r$-process yield 
in accounting for the observed abundances.
Though simulations of quasi-circular NS--NS mergers using Newtonian or conformally flat gravity
have found suitable ejecta, they seem to be in tension with fully general-relativistic 
results which find negligible amounts of ejecta~\citep{faber_review}. This is {\em arguably} because
of strong-field GR effects, such as BH formation and the existence of innermost stable orbits.
As we show, dynamical capture mergers are more promising sources of ejecta, presumably as the stars
are less bound when disruption occurs.

In the remainder of this Letter, we outline our methods for simulating NS--NS mergers
with GRHD, discuss the merger dynamics for a range of impact parameters and 
three different EOSs, and comment on potential GW and EM counterparts.  We find that, while the 
GW signals from these mergers may be challenging to detect with upcoming ground-based detectors, 
they have the potential to source numerous EM transients. Non-merging close encounters can
induce tidal deformations strong enough to crack the NSs' crusts; a merger where the
total mass is above the maximum mass of a single NS can either promptly collapse to a BH or produce a hot, 
rapidly rotating HMNS, where the latter outcome tends to have more massive disks and ejected material.

%--------------------------------------------------------------------------
% Numerical approach
%--------------------------------------------------------------------------
\section{Numerical approach}
We numerically solve the Einstein equations, discretized with finite differences, in the generalized harmonic formulation. 
The hydrodynamics are evolved in a conservative formulation using high-resolution shock-capturing techniques.
Details are given in~\cite{code_paper}.

\begin{figure}
\begin{center}
\hspace{-0.5cm}
\includegraphics[height=2.5in,clip=true,draft=false]{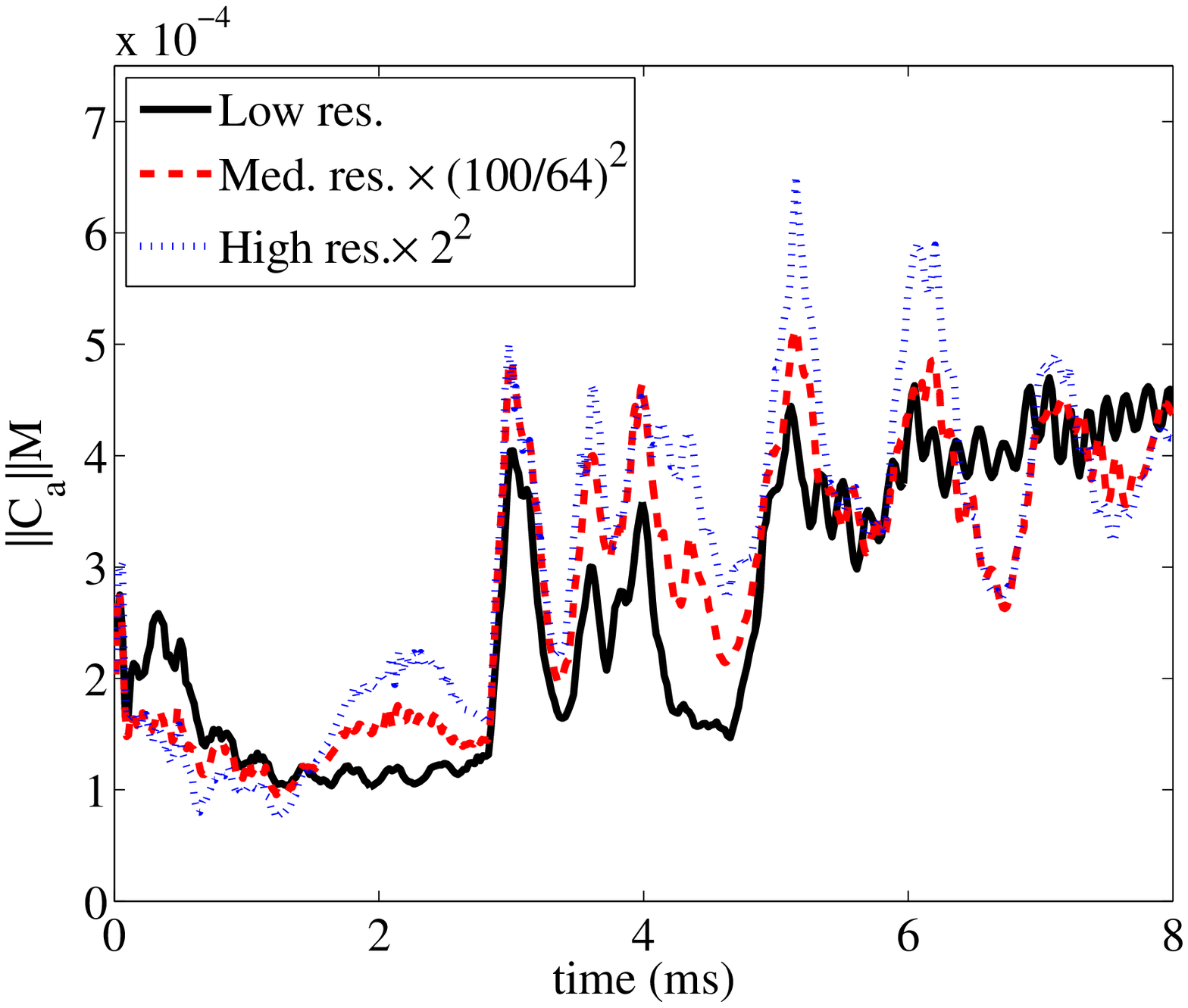}
\includegraphics[height=2.5in,clip=true,draft=false]{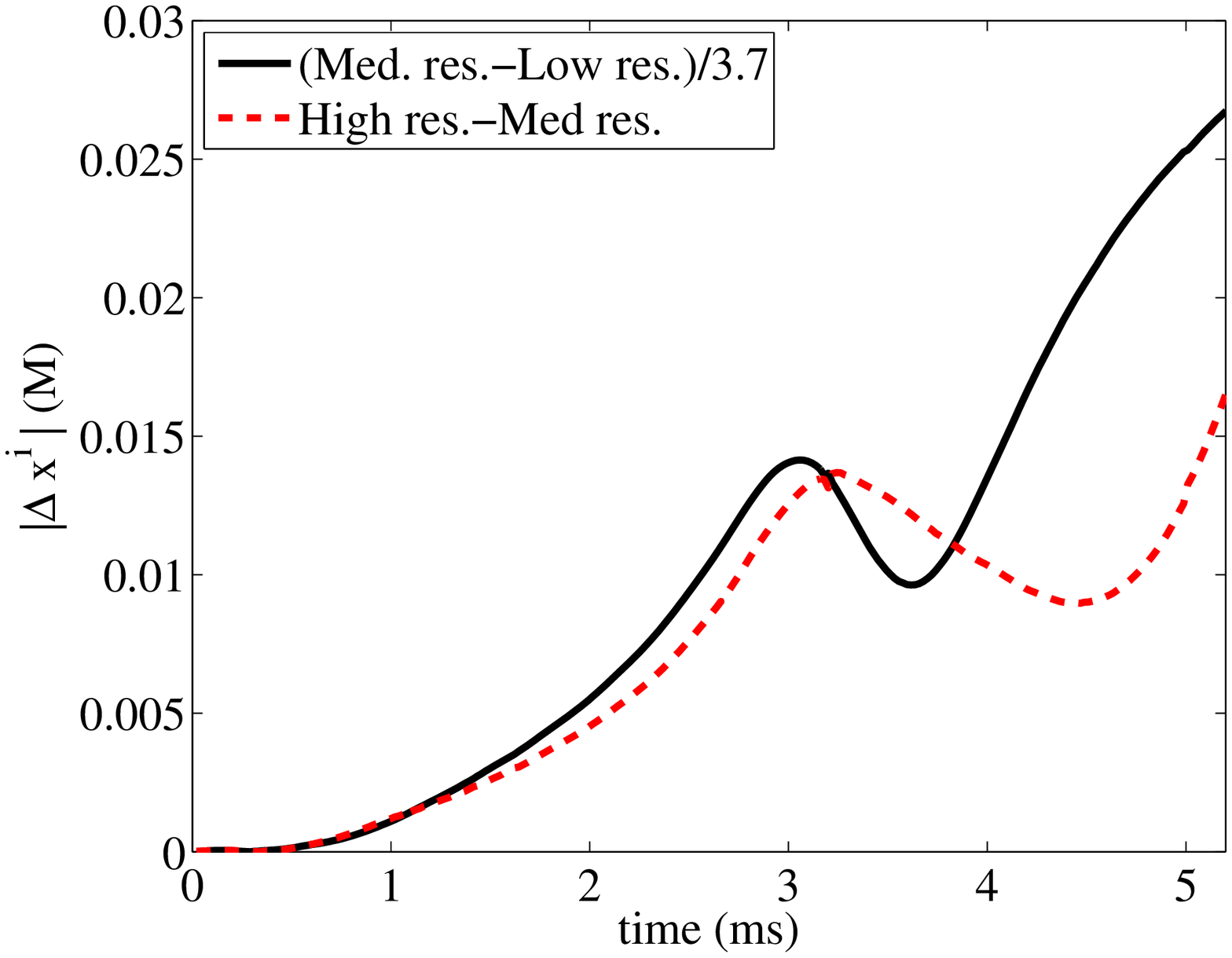}
\caption{
Top: $L^2$-norm of the constraints ($C_a\equiv H_a-\Box x_a$) %in units of $1/M$ 
in the $100M\times100M$ region around the center of mass in the equatorial plane
for the $r_p=7.5$, HB case, with the three resolutions scaled assuming second-order convergence.
Bottom: the difference in the NS center of mass as a function of time from the different
resolution runs for the $r_p=10$, HB case, scaled assuming second-order convergence.
}
\label{conv_plot}
\end{center}
\end{figure}

We use adaptive mesh refinement with up to seven levels that are dynamically adjusted according to truncation
error (TE) estimates. To measure
convergence and TE, we perform a select number of simulations at three
different resolutions.  The low, medium, and high resolutions, respectively, have base levels covered by
$129^3$, $201^3$, and $257^3$ points (with the maximum TE threshold adjusted accordingly), and approximately 64, 100, and 128 points across the
diameter of the NSs on the finest level at the initial time (for the HB EOS).  In Figure \ref{conv_plot},
we show an example of convergence of NS trajectories as well as the constraints of the field equations.
All simulations are performed at medium resolution and results quoted below are from
this resolution, with Richardson extrapolated values given in parenthesis (indicating the quantity's TE)
where multiple resolution data are available.

We use the same gauge, slope limiters, and flux methods as in~\cite{bhns_astro_paper}.
We use the piece-wise polytropic EOS models
labeled 2H, HB, and B from~\cite{read} and include a thermal component 
$P_{\rm th}=(\Gamma_{\rm th}-1)\epsilon_{\rm th}\rho$ with $\Gamma_{\rm th}=1.5$.
These EOSs were designed to span the range of possible EOSs. 
The 2H, HB, and B EOSs, respectively, give NSs with compactions $M_{\rm NS}/R_{\rm NS}$ of 0.13, 0.17, and 0.18 for
$M_{\rm NS}=1.35$ $M_{\odot}$ and maximum masses of 2.83, 2.12, and 2.0 $M_{\odot}$
(unless otherwise stated we use geometric units with $G=c=1$).

We construct initial data by solving the constraint equations in the 
conformal thin-sandwich formulation as described in~\cite{idsolve_paper}.  We
begin the two NSs at a separation of $d=50M$, where
$M=2.7$ $M_{\odot}$ is the total mass of the system (hence $d=200$ km),
and consider various initial velocities which 
we label by $r_p$, the periastron distances of parabolic Newtonian
orbits with the same velocities 
(which will be different from the actual periastron distance of the simulated binaries).  
%(which will be different from the actual periastron distance of the simulated binaries, several of which merge on the first encounter).  
We performed the majority of the simulations using the 
middle compaction HB EOS but ran select impact parameters using all three EOSs.
For all simulations except one we used NSs that both have 
a mass of 1.35 $M_{\odot}$; the other case has a mass ratio of $q=0.8$
and a total mass of 2.88 $M_{\odot}$. We leave a more detailed study of mass-ratio dependence
to future work.

%--------------------------------------------------------------------------
% Results and discussion
%--------------------------------------------------------------------------
\section{Results and discussion}
\subsection{Effect of Impact Parameter}
\begin{figure}
\begin{center}
\hspace{-0.5cm}
\includegraphics[height=2.5in,clip=true,draft=false]{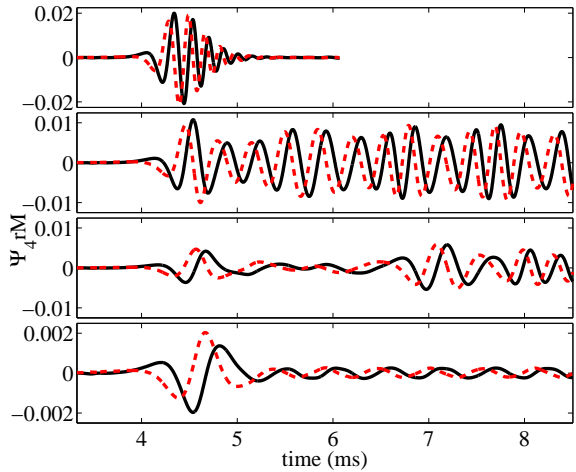}
\caption{ 
Real (solid black) and imaginary (dotted red) components of the Newman--Penrose 
scalar $\Psi_4$ on the axis orthogonal to the orbit measured at $r=100M$ for the $r_p=5$, 7.5, 8.75, and 10 cases with HB EOS.  
}
\label{gwave}
\end{center}
\end{figure}

Using the HB EOS, we consider a range of impact parameters from $r_p/M=2.5$ (we henceforth
quote $r_p$ in units of $M$) to $r_p=20$ (i.e., 10 to 80 km).   
This is well within the range to form a bound system as
\citet{pm63} indicate that for equal masses, GW capture occurs
for $r_p\lesssim1.8/w^{4/7}$ where $w$ is the velocity at infinity. 
(Tidal energy loss~\citep{PressTeuk}, because of the relative scalings 
with distance, is subdominant in determining {\em capture}.)
Binaries that approach with small impact parameters ($r_p=2.5$ and 5) promptly merge and collapse to a BH.
For $r_p=2.5$ the mass and dimensionless spin of the final BH is $M_{\rm BH}/M=0.998(0.995)$ and $a=0.537(0.538)$
while the energy and angular momentum in GWs is $E_{\rm GW}/M=3.7(4.0)\times10^{-3}$ and $J_{\rm GW}/M^2=2.60(2.75)\times10^{-2}$.
For $r_p=5$, $M_{\rm BH}/M=0.985$, $a=0.719$, $E_{\rm GW}/M=1.06\times10^{-2}$, and  $J_{\rm GW}/M^2=6.74\times10^{-2}$; 
Figure \ref{gwave} shows the corresponding GW signals.
For both these cases, the amount of material leftover after merger is $\lesssim10^{-6}$ the original rest mass of the NSs.
The dearth of matter post-merger, and the fact that most of the power of the GW signal is at a relatively high frequency ($\sim5$ kHz),
makes these scenarios less promising sources of observable EM or gravitational radiation.

\begin{figure*}
\begin{center}
\includegraphics[height = 1.5 in]{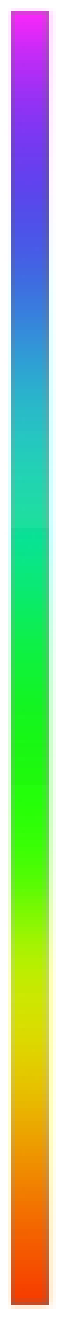}
\put(2,100){$50$ MeV/$m_{\rm n}$}
\put(2,0){$0.05$}
\hspace{1.0 in}
\includegraphics[clip=true,draft=false, viewport=50 10 501 461, width=1.5in]{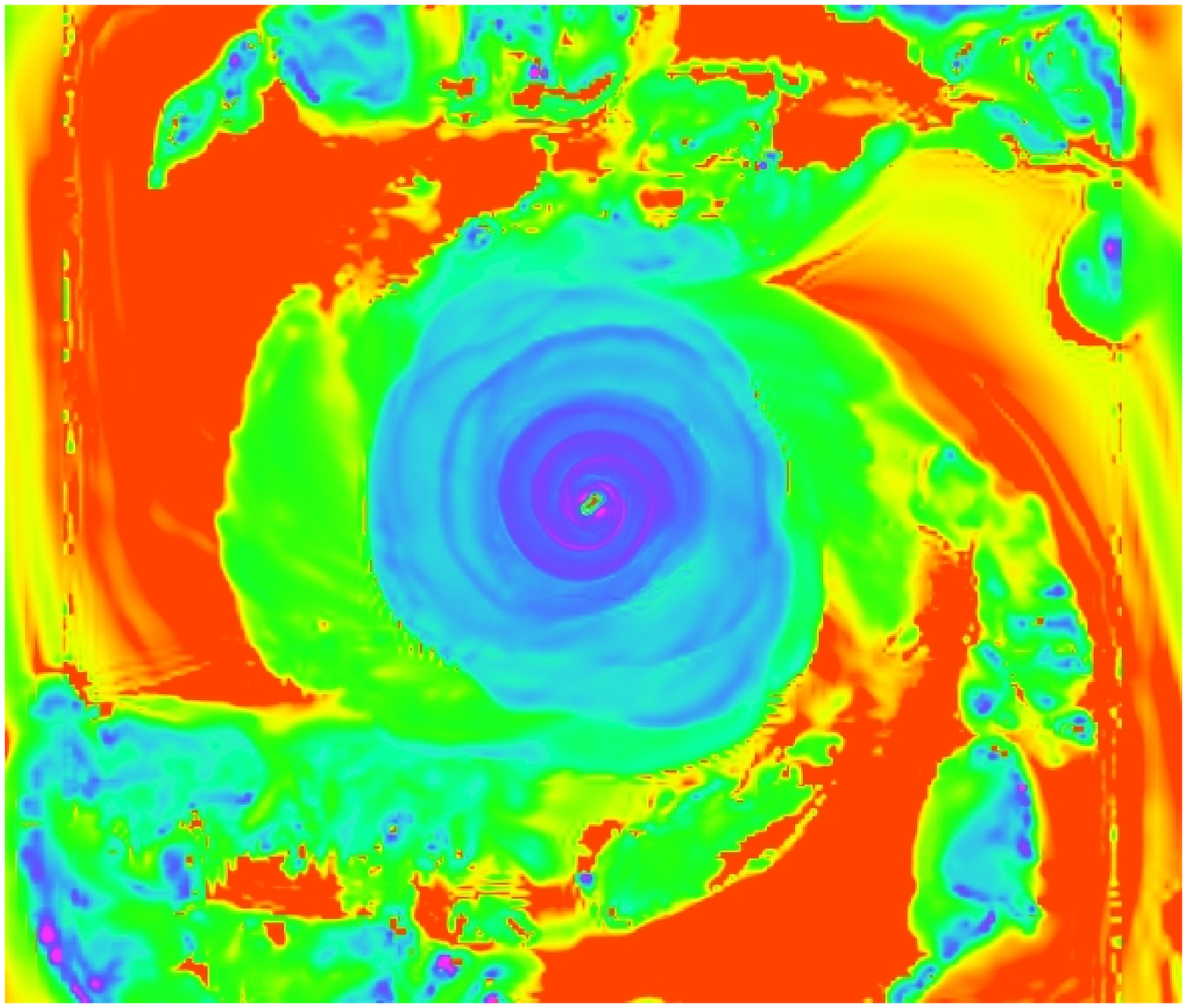}
\includegraphics[clip=true,draft=false, viewport=50 10 501 461, width=1.5in]{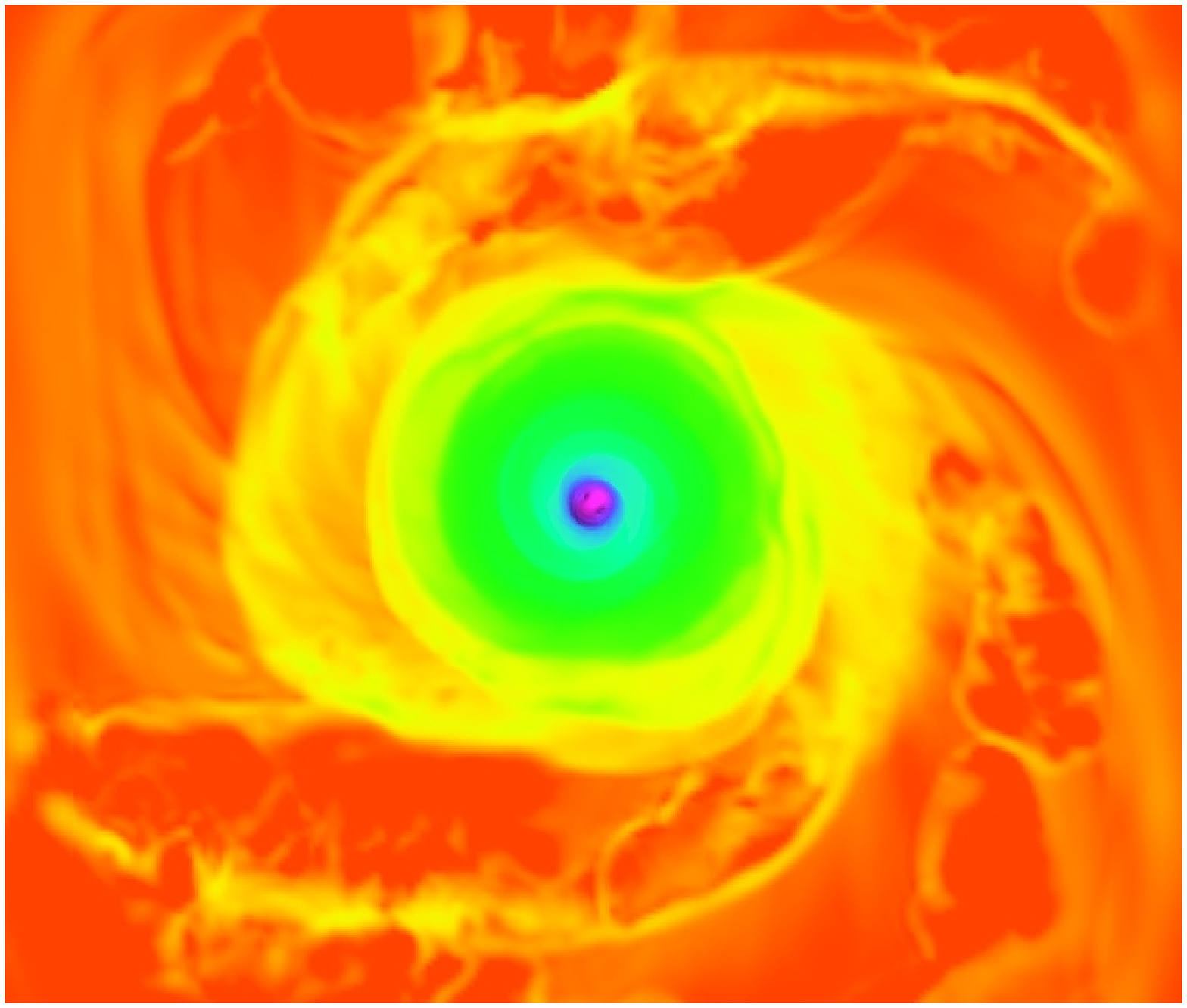}
\includegraphics[clip=true,draft=false, viewport=50 10 501 461, width=1.5in]{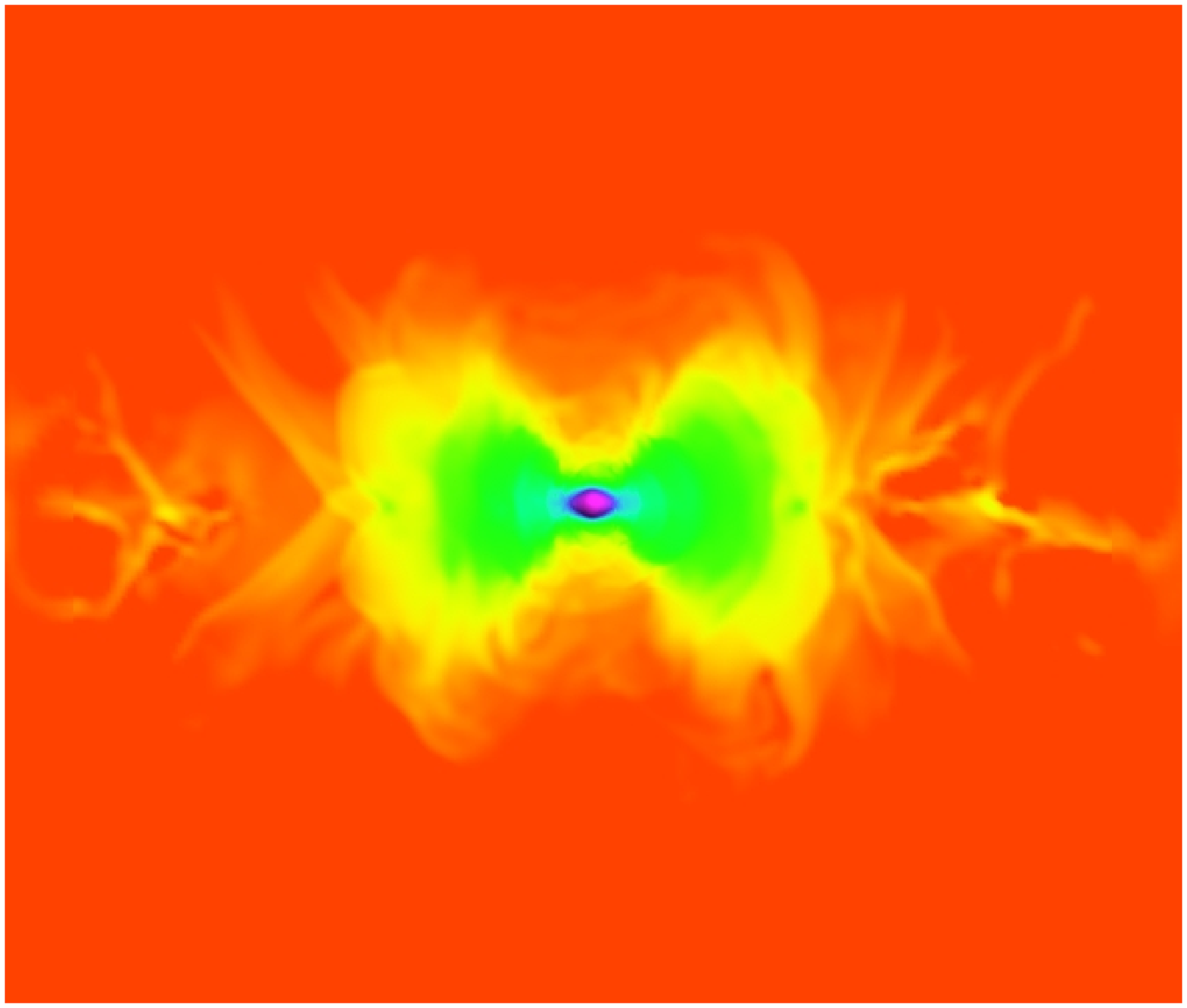}\\
\vspace{0.02in}
\includegraphics[height = 1.5 in]{vertical_scale.eps}
\put(2,100){$10^{15}$ gm cm$^{-3}$}
\put(2,0){$10^{8}$}
\hspace{1.0 in}
\includegraphics[clip=true,draft=false, viewport=50 10 501 461, width=1.5in]{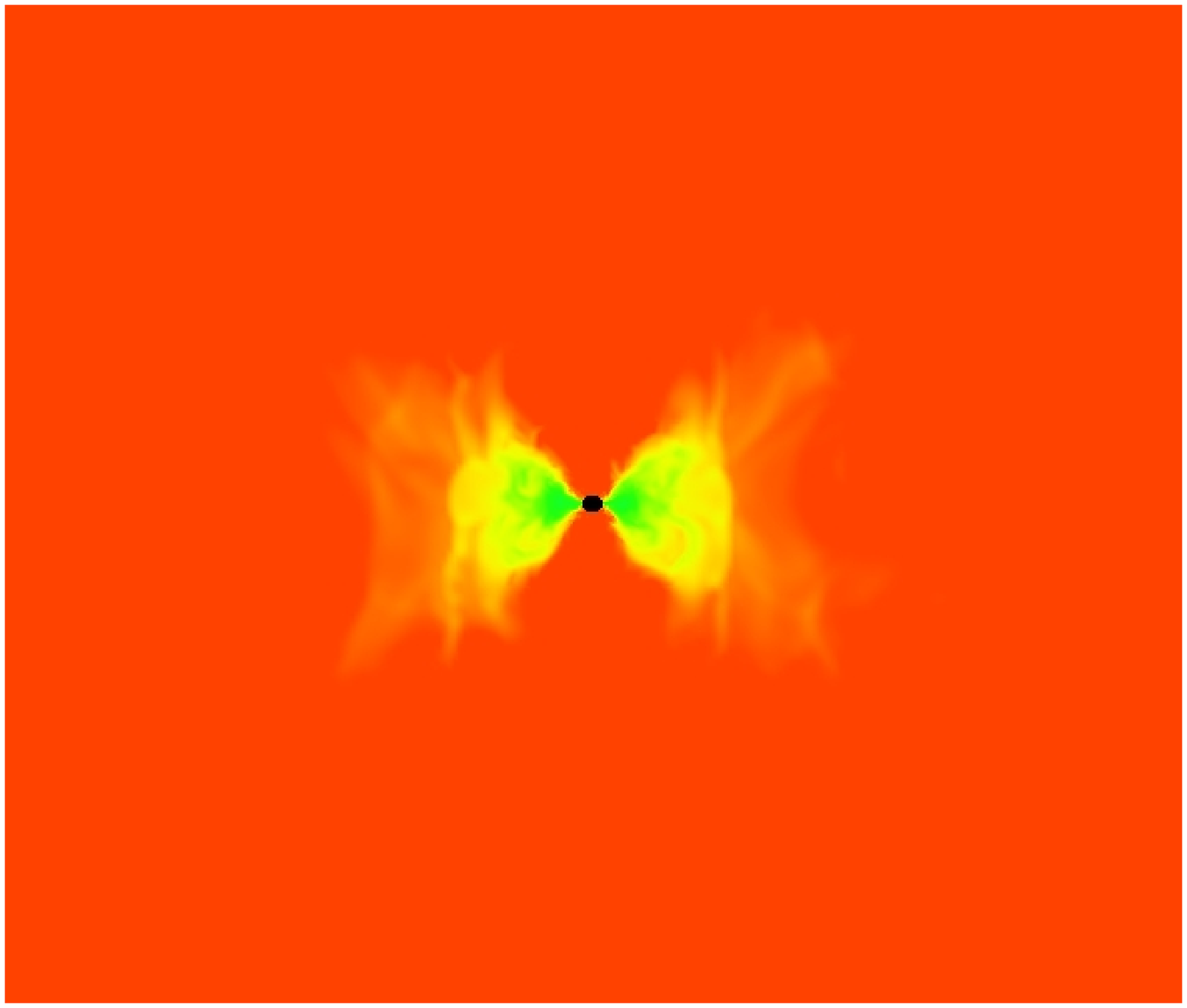}
\includegraphics[clip=true,draft=false, viewport=40 0 511 471, width=1.5in]{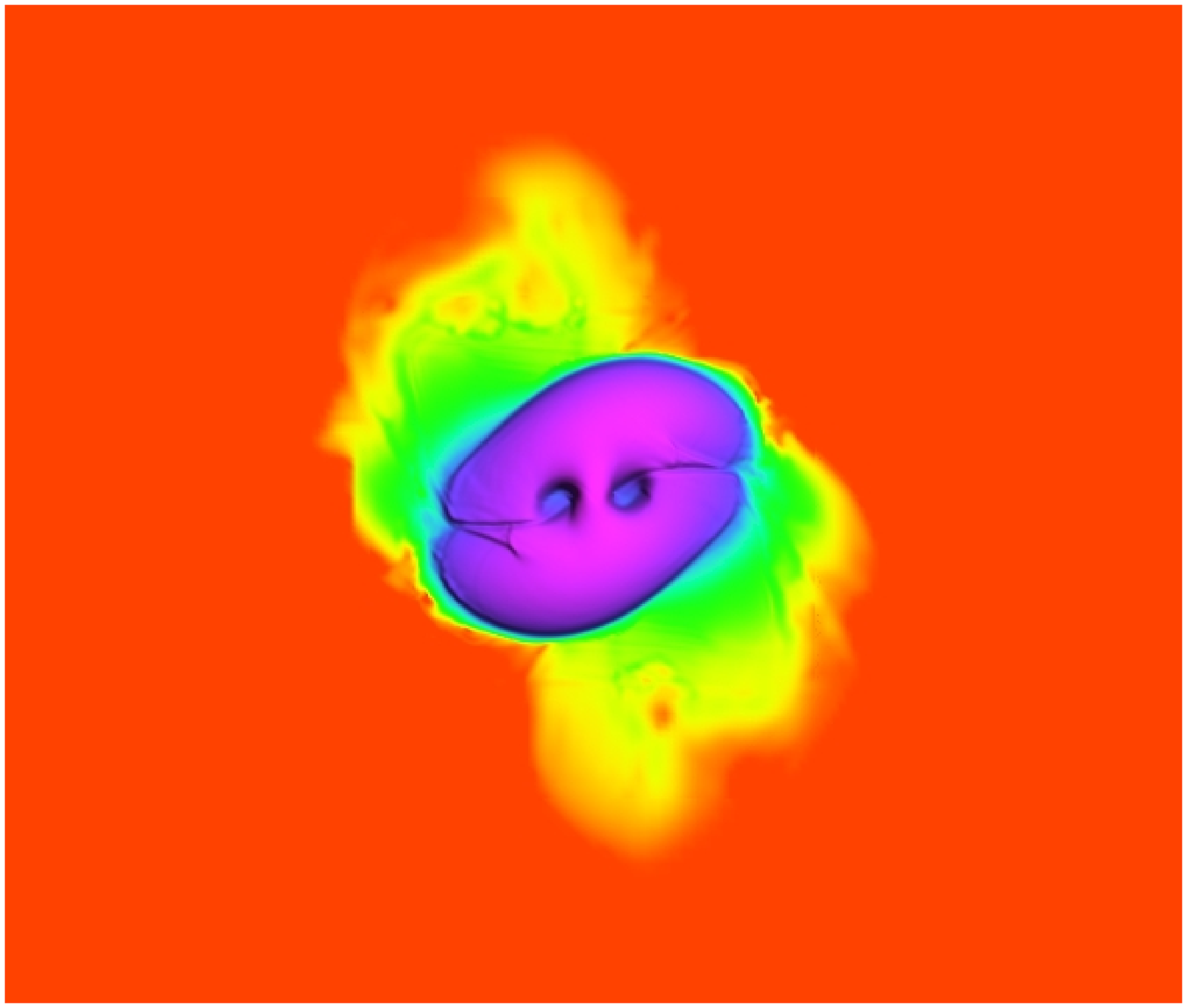}
\includegraphics[clip=true,draft=false, viewport=40 0 511 471, width=1.5in]{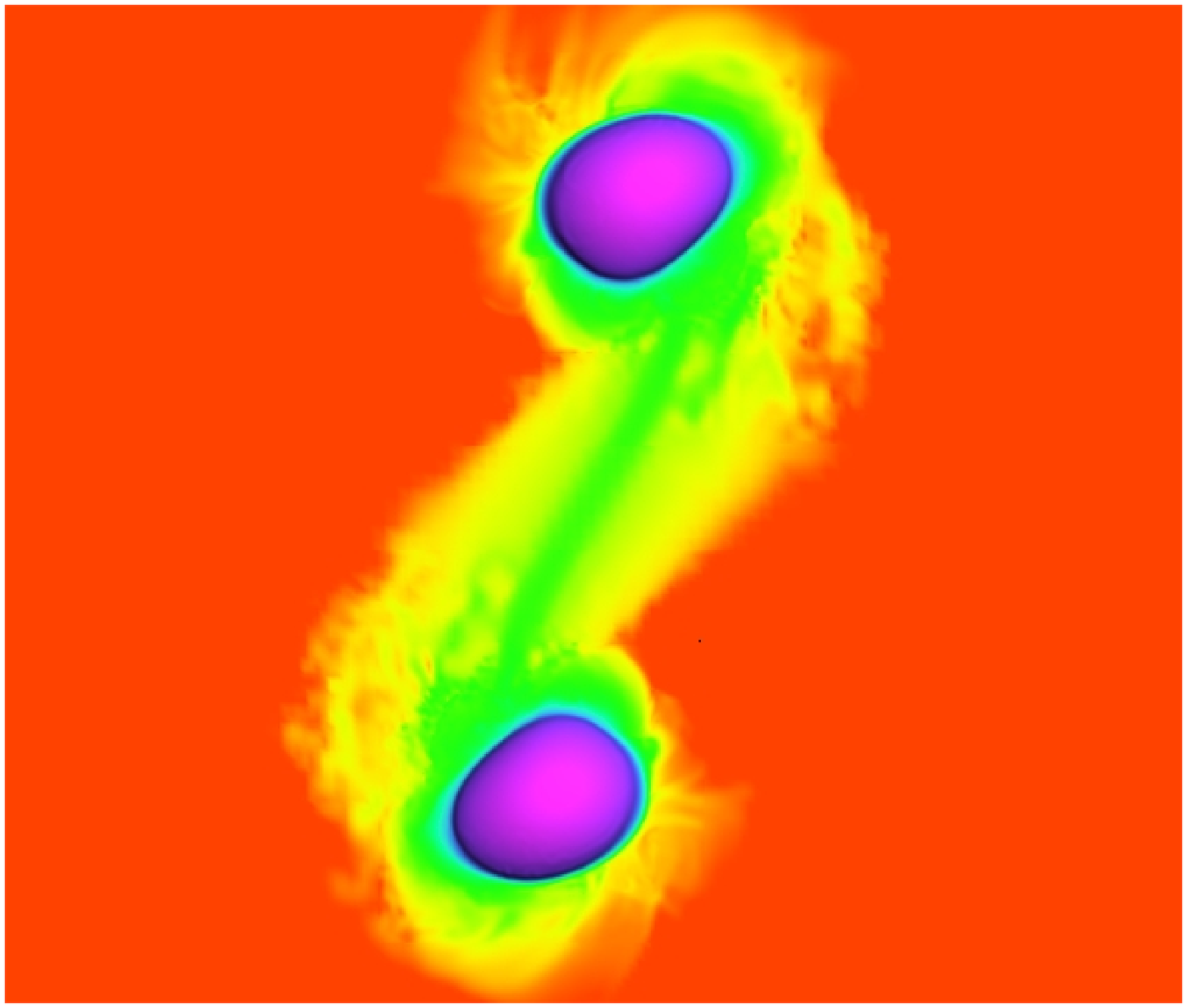}
\caption{
Snapshots of thermal specific energy (top left panel) with a logarithmic color scale 
from $0.05$ to $50$ MeV/$m_{\rm n}$ 
and rest-mass density (other five panels) from $10^{8}$ to $10^{15}$ gm cm$^{-3}$.
The top left and top middle, and bottom middle and bottom right panels show the equatorial
plane, while the other two show a perpendicular plane through the center of mass.
The top panels show an HMNS with surrounding disk and unbound material from the $r_p=8.75$, HB EOS case at $t=13.3$ ms.
The bottom panels show, from left to right,
a BH and a surrounding disk for the $r_p=7.5$ B EOS case at $t=10.2$ ms;
a merger from the $r_p=7.5$, HB EOS case at $t=3.2$ ms; 
NSs with excited $f$-mode perturbations post close encounter from the $r_p=10$, HB EOS case at $t=4.1$ ms.
The first four panels have the same distance scale, where the coordinate 
radius of the HMNS and BH are $\approx13$ and $\approx6$ km, respectively.
The last two panels share a second distance scale; the coordinate separation between
the NSs in the last panel is $\approx73$ km.
}
\label{snapshots}
\end{center}
\end{figure*}

Binaries with larger impact parameters ($r_p=10$, 15, and 20) result in non-merging close encounters followed by 
long elliptic orbits which we did not follow in their entirety due to limited computational resources.  
The close encounters result in pulses of GW radiation and excite $f$-mode oscillations in the stars,
which are also evident in the GW signal (see Figures \ref{gwave} and~\ref{snapshots}).
This $f$-mode excitation was studied in detail in~\cite{gold}
and also found in similar BH--NS encounters~\citep{ebhns_letter,bhns_astro_paper}.  
The induced tidal ellipticity in the $r_p=10$ case 
is greater than the $\delta R/R\approx0.1$ value required to
induce a strain of $u_{\rm strain}\simeq0.1$ and shatter the NS crust (\citet{PhysRevLett.102.191102}; though we are not modeling the crust here).
The energy and angular momentum radiated
in the $r_p=10$ close encounter is $E_{\rm GW}/M=1.472(1.474)\times10^{-3}$ and $J_{\rm GW}/M^2=3.545(3.546)\times10^{-2}$;
for $r_p=15$ and 20 $(E_{\rm GW}/M,J_{\rm GW}/M^2)=(1.64\times10^{-4},8.69\times10^{-3})$ and $(3.8\times10^{-5},2.8\times10^{-3})$,
respectively. Taking this as orbital energy and angular momentum loss gives a Newtonian estimate for the time to the next
close encounter of 65 ms for the $r_p=10$ case. 
For the next largest impact parameter simulated, $r_p=15$, the tidal deformation 
is negligible, and the estimated time to the next close encounter is 1.8 s.
This suggests precursor EM transients associated with crust shattering
for these systems could be produced of order hundreds of milliseconds, but probably 
not more than a few seconds, before merger.

For the intermediate cases ($r_p=7.5$ and 8.75), the stars come into contact and form a single object.  For $r_p=7.5$,
this happens at the first close encounter; for $r_p=8.75$, the stars briefly fly apart before merging.
When the stars come into contact they undergo shock heating and develop features similar to the 
Kelvin--Helmholtz vortices observed in \citeauthor{2012arXiv1204.6240R} (\citeyear{2012arXiv1204.6240R}; see Figure \ref{snapshots}).
Though the total mass is above the maximum for a cold, static star with this EOS, the stars are highly
spun-up and have a significant thermal component ($22\%$--$25\%$) to their internal energy (see Table~\ref{hmns_table}).  
In the vicinity of these objects the (density-weighted) average thermal specific energy is $\epsilon_{\rm th}\approx10$--$20$ MeV/$m_{n}$
where $m_{n}$ is the neutron mass.

These HMNSs produce quasi-periodic GWs with frequency $\sim3.2$ kHz (see Figure \ref{gwave}).  
Though these hypermassive configurations survive the duration of our simulations ($\approx13$ ms), they 
presumably will eventually collapse to form BHs.  In Table~\ref{hmns_table}, we indicate the rate of energy and angular 
momentum loss to GWs at the end of the simulation. From this one can roughly estimate the time it will take for the HMNS to radiate
its remaining angular momentum to GWs assuming a constant $d{J}_{\rm GW}/dt$ (e.g., for the $r_p=7.5$, HB case, it will take $\sim70$ ms). 
However, magnetohydrodynamic effects, such as 
the magnetorational instability,
as well as cooling by neutrino emission,
none of which we take into account, 
will also be important in determining when these stars collapse.  Table~\ref{hmns_table} also lists data from
the unequal mass ratio ($q=0.8$), $r_p=7.5$ case, which shows qualitatively similar behavior to the equal mass case.

\begin{table*}
\begin{center}
{\scriptsize
\begin{tabular}{l l l l l l l l l l}
\hline 
$r_p$ &
EOS &
$J_{\rm tot}/M^2$\ \tablenotemark{a}&
$\left< \epsilon_{\rm th} \right>$\tablenotemark{b} &
$E_{\rm th}/E_{\rm int}$ \tablenotemark{c} &
$M_{0,u}$ \tablenotemark{d} &
$\frac{E_{\rm GW}}{M}\times100 $ \tablenotemark{e} &
$ \frac{J_{\rm GW}}{M^2}\times100$ \tablenotemark{f} &
$ d{E}_{\rm GW}/dt$  \tablenotemark{g} &
$d{J}_{\rm GW}/dt/M$ \tablenotemark{h} \\
\hline
7.5            & HB   & 0.96 &  20   & 0.22    & 0.64 
&3.78(3.91) &  30.5(31.2)   & $1.56\times10^{-5}$    &   $1.23\times10^{-4}$      \\
7.5  ($q=0.8$) & HB   & 0.96 &  14   & 0.17    & 0.57   
&3.36 &  27.5   & $7.60\times10^{-6}$    &   $6.22\times10^{-5}$      \\
7.5            & 2H   & 0.95 &  14   & 0.31    & 4.39
&    0.70 &  10.8   & $3.45\times10^{-7}$    &  $3.20\times10^{-6}$       \\ 
8.75           & HB   & 1.05 &  17   & 0.25    & 2.65
&    2.07 & 24.0    & $1.50\times10^{-5}$     &  $1.14\times10^{-4}$       \\
10           & 2H   & 1.11 &  11   & 0.27    &   6.65
&    0.50 & 9.28    & $2.70\times10^{-6}$     &  $3.48\times10^{-5}$       \\
\hline 

\end{tabular}
\caption{Properties of Hypermassive NS Cases, Measured at $t\approx13.3$ms \label{hmns_table}}

\tablenotetext{1}{Global angular momentum.}
\tablenotetext{2}{Density-weighted average of the thermal component of the specific energy in units of MeV/$m_n$. }% at $t=13.3$ ms}
\tablenotetext{3}{Fraction of internal (Eulerian) energy that is thermal.}
\tablenotetext{4}{Rest mass that is unbound in percent of $M_{\odot}$. }
\tablenotetext{5}{The total energy emitted in GWs through the $r=100 M$ surface.}
\tablenotetext{6}{The total angular momentum emitted in GWs.}
\tablenotetext{7}{Average GW flux of energy.}
\tablenotetext{8}{Average GW flux of angular momentum.}
}
\end{center}

\end{table*}

\subsection{Effect of Equation of State}
In addition to the HB EOS, we also simulated an intermediate impact parameter $r_p=7.5$ using the 
B (softer) and 2H (stiffer) EOSs. For the B EOS, a BH forms soon 
after merger with $M_{\rm BH}/M=0.988$ and $a=0.766$; the total radiated energy and angular
momentum are $E_{\rm GW}/M=3.55\times10^{-2}$ and $J_{\rm GW}/M^2=0.239$, respectively.  For the 2H EOS, 
the total mass is below the maximum for a stable cold NS and the stars fly apart after 
the first collision before eventually settling down to a single massive star. 

We also performed simulations with the B and 2H EOSs and $r_p=10$.  With the softer B EOS, the NSs undergo
a close encounter that is qualitatively similar to the HB EOS.  However, because of the greater
compactness of the NS, the amplitude of the resulting GW pulse is larger with $19\%$ and $13\%$ more energy
and angular momentum, respectively.  Due to the large eccentricity, this will have a significant effect on the subsequent orbit of the NSs.
The estimate of the time to the next close encounter is 50 ms with the B
compared to 65 ms for the HB EOS. Binaries with impact parameters in this range may undergo multiple
close encounters before merging, with the time between the GW bursts
a sensitive function of (in this example) the EOS. 
For the 2H EOS, the stars have larger radii and
graze during the close encounter, merging to a massive star soon after.

\subsection{Possible Post-merger Transients}

\begin{figure}
\begin{center}
\hspace{-0.5cm}
\includegraphics[height=2.5in,clip=true,draft=false]{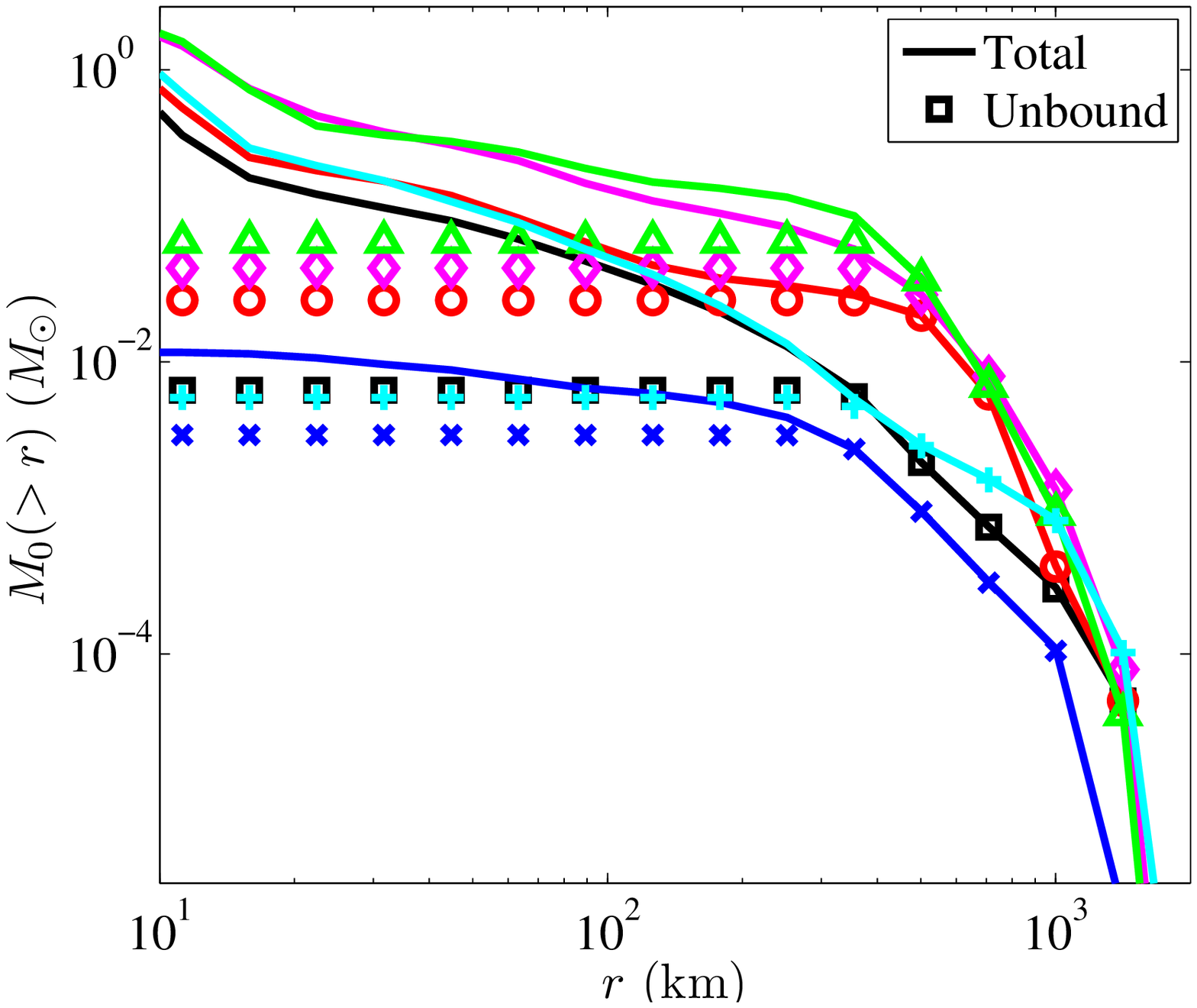}
\includegraphics[height=2.5in,clip=true,draft=false]{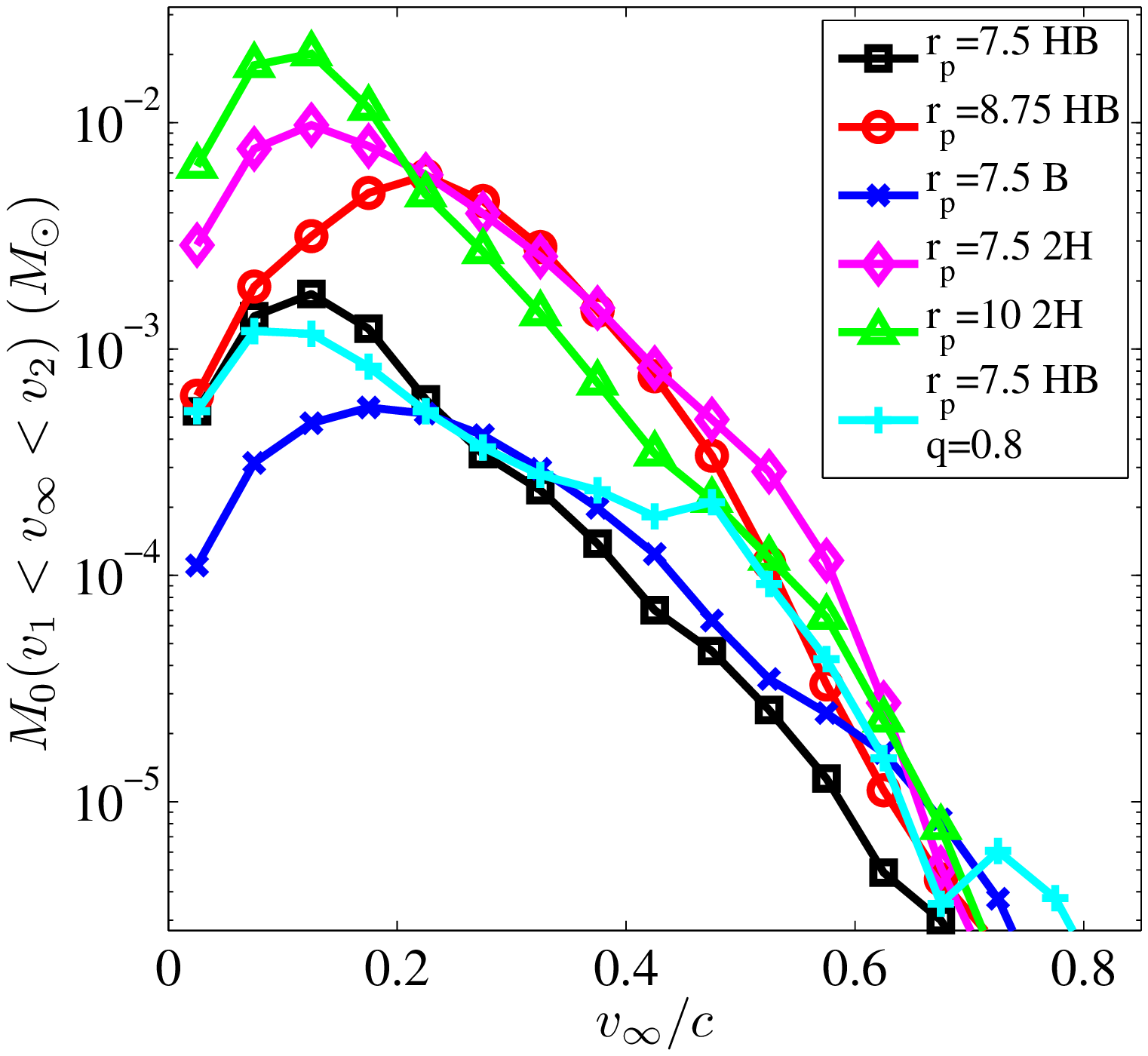}
\caption{
Top: the total and unbound rest mass outside a given radius from the center of mass 
for various cases at $t\approx13$ ms.
Bottom: unbound rest mass with asymptotic velocity grouped in $0.05c$ bins.  
The legend applies to the top panel as well. By this time, the $r_p=7.5$, B case
has collapsed to a BH, the $r_p=7.5$, 2H case is an NS below the maximum mass for this EOS,
while the rest are HMNSs. 
}
\label{matter_plots}
\end{center}
\end{figure}

Intermediate impact parameters ($r_p=7.5$ for the various EOSs and mass ratios, $r_p=8.75$ HB, and $r_p=10$ 2H 
EOS) form HMNSs with non-negligible accretion disks, thought to be necessary for an SGRB progenitor,
and unbound material which could potentially power other EM 
transients (as would presumably a subset of larger impact parameter systems were we
to follow them through merger).  
Figure \ref{matter_plots} shows the amount of matter, total and unbound (fluid cells with outward radial velocity 
and four-velocity time component $u_t<-1$; see also Table~\ref{hmns_table}),
outside a given radius from the 
center of mass and the velocity distribution of the unbound matter.  
The various cases have $0.005$--$0.07M_{\odot}$ unbound material, and roughly $2$--$3$ times more
in a disk. 
As expected, cases with less compact NSs have more unbound material compared to more compact cases. 
Larger impact parameters (which have more angular momentum) also have comparatively 
more unbound material with the most occurring in cases 
where the NSs first come into contact in non-merger close encounters ($r_p=8.75$ HB 
and $r_p=7.5$ and $10$ 2H). 
The unequal mass-ratio merger with $q=0.8$ produces slightly less unbound material than the comparable
equal mass merger.
In all cases, the ejecta is mildly relativistic with asymptotic velocity that peaks in the range $0.1$--$0.3c$.  
This ejecta is presumably neutron rich and will convert to heavy elements
through the $r$-process, the heaviest of which will undergo fission, emitting photons~\citep{Li:1998bw,2005astro.ph.10256K,2010MNRAS.406.2650M}. 
The arguments from~\cite{2010MNRAS.406.2650M} estimate the time scale as
\[t_{\rm peak}\approx0.6\mbox{ d}(M_{\rm u}/3\times10^{-2}M_{\odot})^{1/2}(v/0.2c)^{-1/2}\nonumber\]
with a luminosity, peaking in the optical/near UV, of
\[L\approx4\times10^{42}\mbox{ erg s$^{-1}$}(M_{\rm u}/3\times10^{-2}M_{\odot})^{1/2}(v/0.2c)^{1/2}\nonumber\]
normalized here to the approximate values from the $r_p=8.75$ case. However,
recent calculations using more detailed heavy element opacities suggest that the timescale may be
up to a week with emission peaking in the IR~\citep{kasen_kitp_talk}.

This ejecta is also expected to 
collide with the interstellar medium producing radio waves that will peak on timescales of weeks with brightness~\citep{2011Natur.478...82N}
\begin{eqnarray}
F(\nu_{\rm obs})&\approx&0.4(E_{\rm kin}/2\times10^{51}\mbox{ erg})(n_0/0.1{\rm cm}^{-3})^{7/8}\nonumber\\
&&(v/0.2c)^{11/4}(\nu_{\rm obs}/{\rm GHz})^{-3/4}(d/100{\rm Mpc})^{-2}\mbox{ mJy}\nonumber
\end{eqnarray}
where $n_0$ is the density of the surrounding environment (we use $n_0\sim0.1$ cm$^{-3}$ for GC cores;~\citealt{2012arXiv1204.6240R}), 
$\nu_{\rm obs}$ is the observation frequency, $d$ the distance, and we have 
normalized the kinetic energy and velocity to the $r_p=8.75$, HB simulation. 

%--------------------------------------------------------------------------
% Conclusions
%--------------------------------------------------------------------------
\section{Conclusions}
We have performed GRHD simulations modeling dynamical capture NS--NS mergers,
giving direct estimates of the corresponding GW emission and merger outcome varying
impact parameter, EOS, and mass ratio. 
By measuring pre-merger tidal deformation and post-merger stripped material (bound and ejected), we
have also speculated on related EM transients. 

Regarding transients that may precede the merger, 
non-merging close encounters can lead to tidal
deformations strong enough to crack the NSs' crust and tap into the $\sim10^{46}$ erg
stored in elastic energy~\citep{1995MNRAS.275..255T}, potentially causing flaring activity
from milliseconds up to possibly a few seconds before merger. Though a different mechanism
and time scales, the signature
could be similar to resonance induced cracking for quasi-circular inspirals
proposed in~\cite{PhysRevLett.108.011102}.
The cracking of the NS crust is one possible explanation for SGRB precursors
observed by \emph{Swift}~\citep{precursors}.  

We find that dynamical capture mergers can result in prompt
BH formation or the formation of an HMNS depending on impact parameter and
EOS.  The HMNSs will be long lived due to their rapid rotation and 
thermal energy, giving them the potential to seed large magnetic fields and source
intense transients during collapse.

In contrast to what was found in general-relativistic
studies of quasi-circular NS--NS mergers, we find that dynamical capture mergers
can result in massive disks even for equal mass binaries, and 
can result in up to a few percent of a solar mass in ejecta.  This mildly relativistic 
ejecta can produce potentially observable optical and radio transients.  
The amount of ejecta found here is similar to the $0.009$--$0.06$ $M_{\odot}$ found with
Newtonian gravity~\citep{2012arXiv1204.6240R}, though not 
for comparable impact parameters ($r_p\leq5$). However, what is qualitatively consistent
with the Newtonian setups  
is that we observe the largest amounts of unbound material for grazing collisions. 

Regarding GW detectability, the high frequency of the merger-ringdown
or quasi-periodic signals from the HMNS will be difficult to observe with
AdLIGO.  Individual bursts
from close encounters would also not be detectable except for very nearby events. For example,
an $r_p=10$, HB EOS merger at $d=100$ Mpc has sky-averaged S/N for AdLIGO of $\approx0.9$.  
This implies that if dynamical capture NS--NS mergers constitute a fraction of SGRB progenitors,
a further subset of these
will not have a detectable GW counterpart.
GW signals from larger $r_p$ binaries undergoing numerous close encounters would have 
larger S/N,
and the timing between bursts will be a sensitive function of
the orbital energy, containing information about the EOS, for example.
We defer a detailed study of GW detectability to future work~\citep{upcoming_ecc_gwave}.

%--------------------------------------------------------------------------
% Acknowledgments.
%--------------------------------------------------------------------------
\acknowledgments
We thank Chris Thompson for useful discussions, as well as participants
of the KITP ``Rattle and Shine'' conference (2012 July).
This research was supported by NSF grant PHY-0745779, and resources from
NSF XSEDE provided by NICS under grant TG-PHY100053
and the Orbital cluster at Princeton University.

%--------------------------------------------------------------------------
% References
%--------------------------------------------------------------------------

\bibliographystyle{hapj}
\bibliography{nsns}

\end{document}